\documentclass[11pt]{article}
\usepackage{amsmath,amssymb}
\usepackage{graphicx}
\usepackage{float}
\newcommand{\Slash}[1]{{\ooalign{\hfil/\hfil\crcr$#1$}}} 

\renewcommand{\title}[1]{%
	\begin{center} \Large \bf #1 \end{center}%
	}
\renewcommand{\author}[3]{%
	\begin{center} #1 #2 \\ %
	  {\small #3 \small}%
	\end{center}%
	}

\begin{document}
\baselineskip 5mm
\title{%
Analysis of the Dynamical Chiral Symmetry Breaking in QCD at Finite Temperature and Density using the Non-Perturbative Renormalization Group
}%
\author{%
}{%
Ken-Ichi Aoki, Daisuke Sato and Masatoshi Yamada
}{%
{\em Institute for Theoretical Physics, Kanazawa University}
}
\begin{abstract}
We analyze the Nambu--Jona-Lasinio model which is a chiral effective model of QCD by using the non-perturbative renormalization group at finite temperature and finite density. We discuss the chiral phase diagram in the model.  We include the large-$N$ non-leading contribution in the beta function and discuss its effects for the phase boundary.
\end{abstract}
\section{Introduction}
~~~~Understanding the non-perturbative Quantum Chromodynamics (QCD) is one of the important subjects in elementary particle physics. The phase diagram of QCD at finite temperature and finite density has been studied intensively. There are various non-perturbative approaches to QCD, such as the lattice QCD, the mean-field approximation (MFA) and the Schwinger-Dyson equations (SDEs). However, these mothods have serious problems respectively. The lattice QCD is a powerful method because of the first-principle caluclation. However, it is difficult to maintain the chiral symmetry on the lattice field. At finite density, the QCD action has a complex phase due to the quark chemical potential. Owing to this problem, called the sign problem, the statistical errors of simulation cannot be controlled easily. The MFA or the SDEs with the ladder approximation have been used in various types of models. However, these methods have difficulties in the further improvement of the approximation without which the strong gauge dependences cannot be cured. \\
\indent We study the non-perturbative properties using the non-perturbative renormalization group (NPRG). The NPRG method gives us not only the equivalent results to MFA and SDEs in the lowest order approximation \cite{boso1}, but also the systematic method for improving approximation. We don't confront the sign problem at finite density. However, the NPRG breaks gauge symmetry due to inclusion of the momentum cutoff and also it has the gauge dependences as MFA and SDEs. These problems can be treated by systematic improvement of approximation. It reported that the gauge dependence has almost been wiped away by including the non-ladder diagrams to the beta functions \cite{sato}.\\
\indent We appliy the NPRG to the Nambu--Jona-Lasinio (NJL) model \cite{nambu-jona} at finite temperature and finite density. The NJL model is the effective theory with four-fermi interactions and describes the S$\chi$SB of QCD. We compare the results improved by the NPRG method with the mean field approximation. 
\section{Non-perturbative Renormalization Group}
~~~~We briefly explain the basic idea of NPRG \cite{WilK} in quantum field theory. We divide the degrees of freedom of quantum field $\phi (p)$ into the higher modes with $|p|>\Lambda$ and the lower modes with $|p| <\Lambda$ in the Euclidean space. Then we define the effective action $S_{\rm eff}[\phi;\Lambda]$, called the Wilsonian effective action, by integrating out only the higher modes in the path integral
\begin{align}\label{wilsonian}
Z=\int ^{\Lambda _0}{\mathcal D}\phi~e^{-S_0}=\int ^{\Lambda}{\mathcal D}\phi_< \int ^{\Lambda _0}_{\Lambda}{\mathcal D}\phi_> ~e^{-S_{0}[\phi_<+\phi_>]} =\int ^{\Lambda}{\mathcal D}\phi_<~e^{-S_{\rm eff}[\phi _<;\Lambda]},
\end{align}
where $S_0$ is the initial (bare) action at the initial cutoff $\Lambda _0$. The NPRG equation describes the dependence of the Wilsonian effective action on the cutoff $\Lambda$,
\begin{align}\label{rge}
\frac{\partial}{\partial \Lambda} S_{\rm eff}[\phi;\Lambda]=\beta[S_{\rm eff};\Lambda].
\end{align}
The right hand side of this equation is called the beta function. It is evaluated as the infinitesimal change of the Wilsonian effective action by infinitesimally lowering the cutoff $\Lambda$.\\
\indent There are various formulations of NPRG equation \cite{WH, Pol,Wet}. In this paper, we adopt the Wetterich flow equation \cite{Wet} which is a differential equation for the Legendre effective action with IR cutoff,
\begin{align}\label{wetterich}
\partial _{\Lambda}\Gamma _{\Lambda}[\Phi]=\frac{1}{2}{\rm STr}\left\{ \left[ \frac{\overrightarrow \delta }{\delta \Phi}\Gamma _{\Lambda}[\Phi]\frac{\overleftarrow \delta }{\delta \Phi}+R_{\Lambda}\right]^{-1}\cdot  (\partial _{\Lambda}R_{\Lambda}) \right\} ,
\end{align}
where $R_{\Lambda}$ is the cutoff profile function which divides the higher and the lower modes of quantum field. This equation is exact. It describes the development of the effective action starting from the bare action $S_0$ to the  Legendre effective action $\Gamma _{\Lambda=0}$. \\
\indent The equation (\ref{wetterich}) is a functional differential equation and we cannot solve it exactly. We have to make some approximation. First, the effective action is expanded into power series of derivative of fields,
\begin{align}\label{de}
\Gamma _{\Lambda}[\phi]=\int d^4x\left[ V_{\Lambda}(\phi)+\frac{1}{2}Z_{\Lambda}(\phi)(\partial _{\mu}\phi)^2+\frac{1}{2}Y_{\Lambda}(\phi)(\partial ^2 \phi)^2+\cdots  \right],
\end{align}
where $V_{\Lambda}$ is the effective potential generated and  $Z_{\Lambda}$ and $Y_{\Lambda}$ are the field renormalization factors. This method is called the derivative expantion. Next, we ignore all the conections to terms with derivatives. Then the effective action is represented by the effective potential $V_{\Lambda}$. This approximation, called the local potential approximation (LPA) \cite{LPA}, allows us to evaluated the effective action only with the zero momentum mode of fields. We reduce Eq.~(\ref{wetterich}) to be a partial differential equation for the effective potential $V_{\Lambda}$. Furthermore if the effective potential is spanned by the polynomials of fields, we get infinitely coupled ordinary differential equations for the expansion coefficients (the coupling constants).
\section{Nambu--Jona-Lasinio model}
\begin{figure}[H]
\begin{center}
\includegraphics[width=160mm]{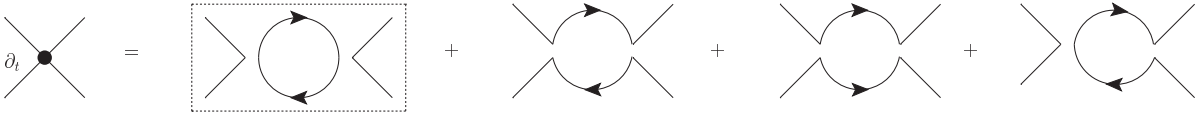}
\end{center}
\caption{Feynman diagrams contributing to the beta function for the four-fermi interactions.}
\label{4fermi}
\end{figure}

~~~~The Lagrangian of the NJL model \cite{nambu-jona} with one flavor and one color is given by 
\begin{align}\label{NJL}
{\mathcal L}_{\rm NJL}= {\bar \psi}i{\Slash \partial}\psi+\frac{G_0}{2}\{ ({\bar \psi}\psi )^2+({\bar \psi}i\gamma _5\psi )^2\} .
\end{align}
This Lagrangian is invariant under the chiral U(1) transformation: $\psi \to e^{i\gamma _5\theta}\psi$. The four-fermi coupling constant $G$ corresponds to the fluctuation of the chiral order parameter: $\langle ({\bar \psi}{\psi})^2 \rangle$, therefore, we may conclude the S$\chi$SB by divergence of the four-fermi coupling constant at a finite energy scale. \\
\indent The NJL model at finite temperature and finite density is defined by the following Euclidean bare action,
\begin{align}\label{NJLb}
S_0=\int _0^{1/T}d\tau \int d^3x\left[ {\bar \psi}{\Slash \partial}\psi  +\mu{\bar \psi}\gamma _0\psi -\frac{G_0}{2}\{ ({\bar \psi}\psi )^2+({\bar \psi}i\gamma _5\psi )^2\}\right].
\end{align}
The four-fermi interaction $G_0$ generates the effective four-fermi coupling constant $G_{\Lambda}$ by quantum corrections, that is, by driving the infrared cutoff scale to the low energy. The effective action in LPA is denoted by
\begin{align}\label{NJLe}
\Gamma _{\Lambda}=\int _0^{1/T}d\tau \int d^3x\left[ {\bar \psi}{\Slash \partial}\psi  +\mu{\bar \psi}\gamma _0\psi -\frac{G_{\Lambda}}{2}\{ ({\bar \psi}\psi )^2+({\bar \psi}i\gamma _5\psi )^2\} \right].
\end{align}
We calculate the generated four-fermi interactions through the diagrams in Fig.~\ref{4fermi}. The first diagram in the dashed box in Fig.~\ref{4fermi} is the so-colled large-$N$ leading term. If we adopt only this term in the beta function of the four-fermi coupling constant, we get the equivarent results to the MFA or the ladder SDEs. Note that we can add the large-$N$ non-leading terms without any difficulty.\\
\indent In this study, we use the 3-$d$ optimized cutoff function \cite{opt} in Eq.~(\ref{wetterich}),
\begin{displaymath}
R_{\Lambda}(\boldsymbol{p})=\boldsymbol{\Slash p}\left( \displaystyle \frac{\Lambda}{\boldsymbol{|p|}}-1\right) \theta (1-\frac{\boldsymbol{p}^2}{\Lambda ^2})=\boldsymbol{\Slash p}~r(\boldsymbol{p}/{\Lambda}).
\end{displaymath}
At finite temperaure, the time direction momentum is discretized and the integration of it changes to the Matsubara summation, therefore, the 3-$d$ cutoff function is appropriate to write down the renormalization group  equations in a simple form. 
\section{Results}
\begin{figure}
 \begin{minipage}{0.5\hsize}
  \begin{center}
   \includegraphics[width=55mm,angle=270]{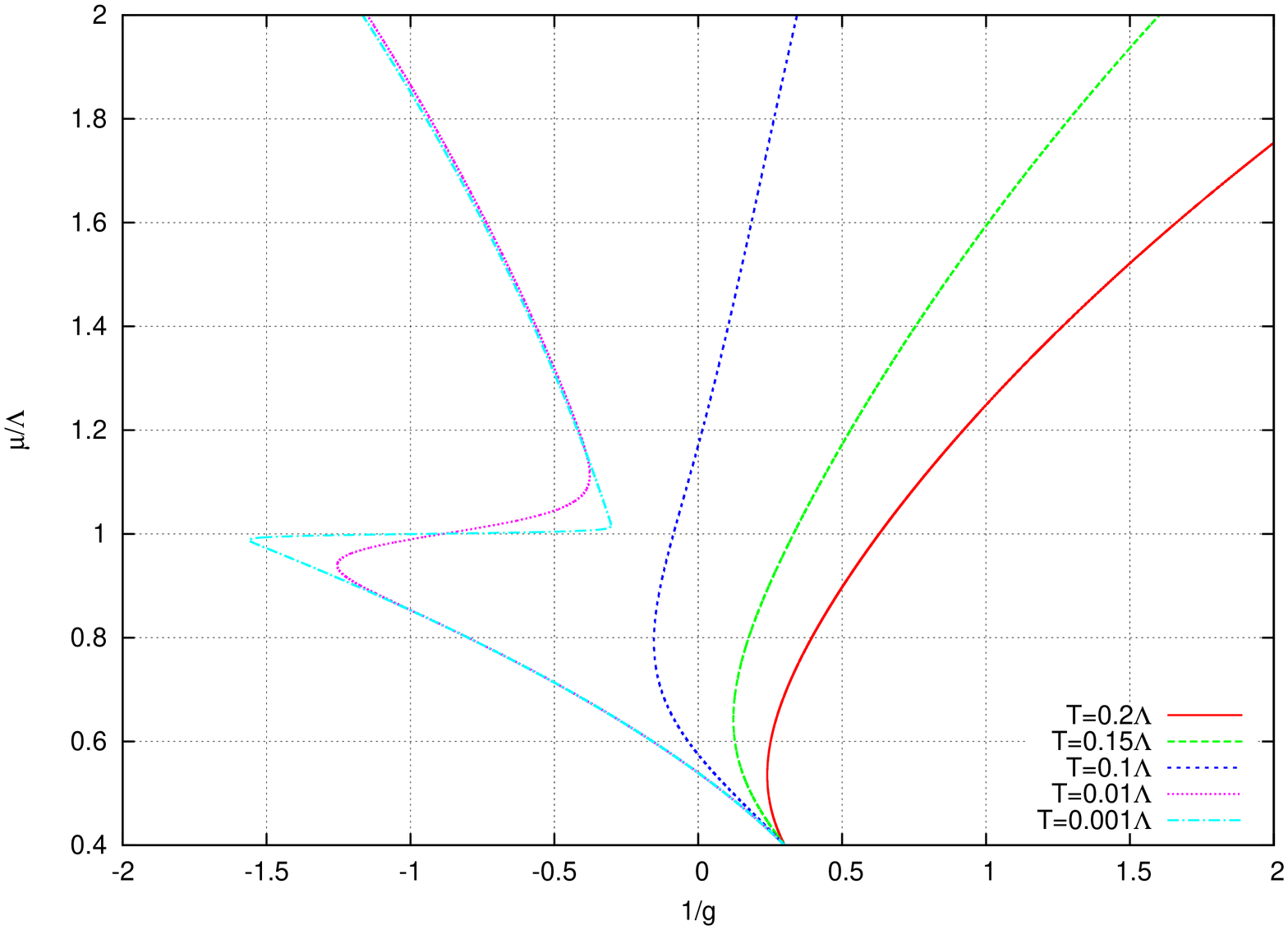}
  \end{center}
  \caption{The RG flows of ${\tilde g}$ of the large-$N$ leading on the ${\tilde g}-{\tilde \mu}$ plane at ${\mu/\Lambda_0 =0.3}$ and various temperatures. We set the initial inverse four-fermi coupling to 0.3.}
  \label{leading}
 \end{minipage}
 \begin{minipage}{0.01\hsize}
\end{minipage}
 \begin{minipage}{0.5\hsize}
  \begin{center}
   \includegraphics[width=55mm, angle=270]{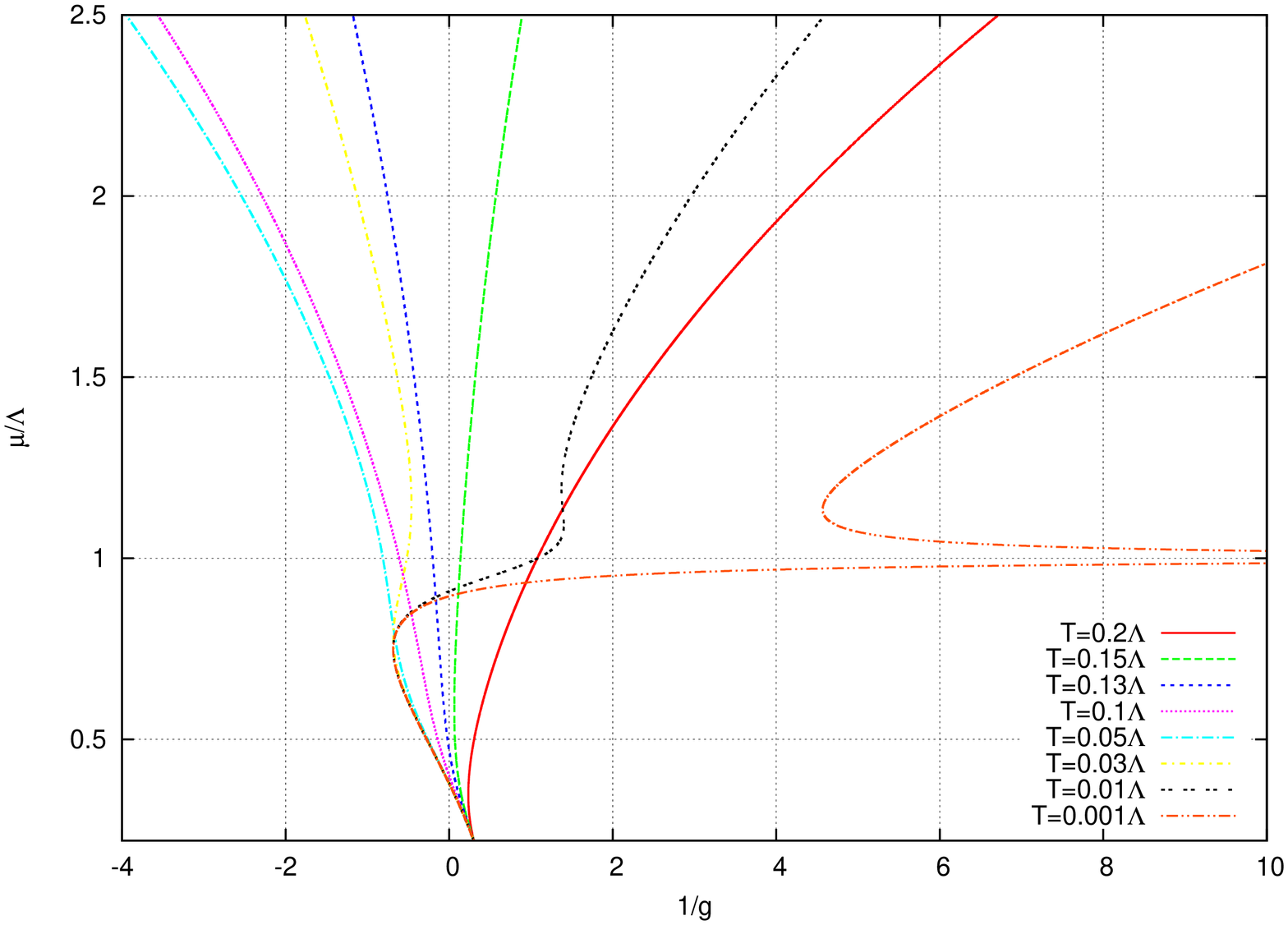}
  \end{center}
  \caption{The RG flows of ${\tilde g}$ of the large-$N$ non-leading  on the ${\tilde g}-{\tilde \mu}$ plane at ${\mu/\Lambda _0 =0.22}$ and various temperatures. We set the initial inverse four-fermi coupling to 0.3.}
  \label{non-leading}
 \end{minipage}
\end{figure}

\begin{figure}
\begin{center}
\includegraphics[width=120mm]{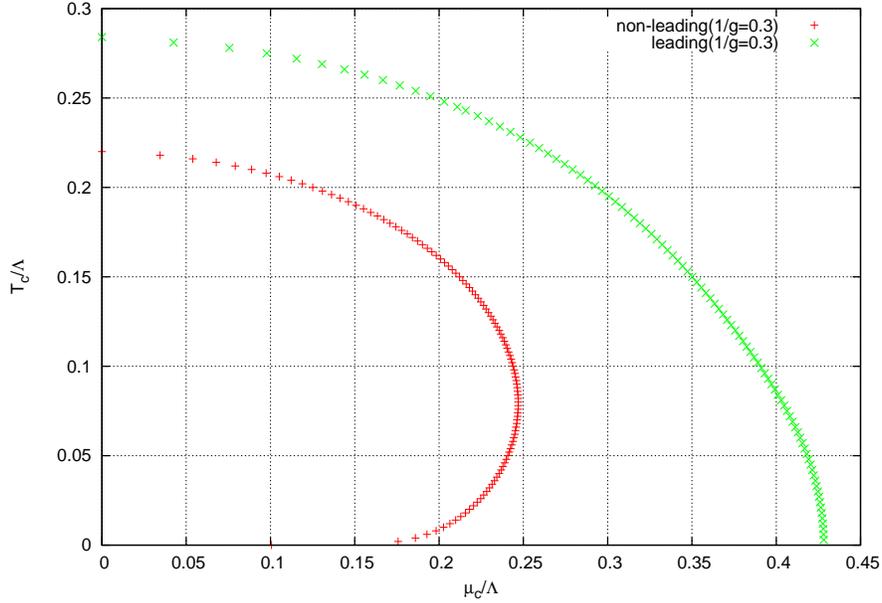}
\end{center}
\caption{The chiral phase diagram in the large-$N$ leading and non-leading calculation}
\label{phase}
\end{figure}

~~~~The RGEs we solve are the following three simultaneous differential equations,
\begin{align}
\nonumber
\partial _t{\tilde g}&=2{\tilde g}-\frac{1}{3}(4I_0-I_1),\\
\label{RGE}
\partial _t {\tilde T}&={\tilde T},\\
\nonumber
\partial _t {\tilde \mu}&={\tilde \mu},
\end{align}
where $\partial _t=-\Lambda\partial _{\Lambda}$, $1/{\tilde  g}=g=G_{\Lambda}\Lambda ^2/4\pi ^2$, ${\tilde T}=T/\Lambda$ and ${\tilde \mu}=\mu/\Lambda$. The threshold functions $I_0({\tilde T},{\tilde \mu})$ and $I_1({\tilde T},{\tilde \mu})$ are given by  the large-$N$ leading diagram and the non-leading diagrams respectively in Fig.~\ref{4fermi} , and are expressed as
\begin{align}
I_0&=\left. \left[ \left( \frac{1}{2}-n_+ \right) +\left( \frac{1}{2}-n_-\right)+\frac{\partial}{\partial \omega}(n_++n_-)\right] \right|_{\omega \to 1},\\
I_1&=\left. \left[ \frac{1}{(1+{\tilde \mu})^2}\left( \frac{1}{2}-n_+ \right) +\frac{1}{(1-{\tilde \mu})^2}\left( \frac{1}{2}-n_-\right)+\frac{1}{1+{\tilde \mu}}\frac{\partial}{\partial \omega}n_+ +\frac{1}{1-{\tilde \mu}}\frac{\partial}{\partial \omega}n_-\right] \right|_{\omega \to 1},
\end{align}
where $n_{\pm}$ are the Fermi-Dirac distribution functions, $n_{\pm}=({e^{(\omega \pm {\tilde \mu})/{\tilde T}}+1})^{-1}$. The large-$N$ non-leading effect $I_1$ contributes towards the restoration of chiral symmetry. Especially, this effect is large at low temperature and high density. If temperature vanishes, the threshold function $I_1$ diverges at ${\tilde \mu}=1 (\mu =\Lambda)$ when the cutoff reaches the fermi surface.\\
\indent We numerically solve the RGEs (\ref{RGE}). We may conclude the S$\chi$SB when ${\tilde g}$ passes the origin. We show the RG flows of ${\tilde g}$ of the large-$N$ leading and the non-leaing in Fig.~\ref{leading} and \ref{non-leading} respectively. We draw RG flows after ${\tilde g}$ passes the origin. This is justified by the "Weak solution" method \cite{Kuma} in case of the large-$N$ leading. Infact, the inverse four-fermi coupling constant ${\tilde g}$ corresponds to the mass squared of meson fields introduced as the auxiliary fields. In other words,  ${\tilde g}$ corresponds to the curvature of the effective potential of meson field  at the origin, and thus, the negative of ${\tilde g}$ means that the curvature of the effective potential at the origin becomes negative. In this case, the effective potential has a global minimum at non-zero expectation value of meson field, then the S$\chi$SB occurrs. If we directly solve the RGE for the four-fermi coupling constant, the RG flows stop on the way because the four-fermi coupling constant diverges at finite $t$. However, as shown in Fig.~\ref{leading}, some flows go to the negative region, and then comes back to the positive region afterwards. We may regard such turn-over flows as symmetric phase in case of the large-$N$ leading. In case of the large-$N$ non-leading, it is difficult to justify of this interpretation. However, we adopt the some criterion here. \\  
\indent The phase diagrams of the large-$N$ leading and the non-leading calculation are shown in Fig.~\ref{phase}. We can see the drastic difference of behavior of phase boundaries due to the large-$N$ non-leading effects at low temperature and high density.
\section{Summary and Discussion}
~~~~We study the Nambu--Jona-Lasinio model by using the non-perturbative renormalization group at finite temperature and finite density. We discuss the difference of the RG flows and phase boundaries between the mean field approximation results and those including the large-$N$ non-leading effects. We find that the large-$N$ non-leading effects become large and contribute to the restoration of chiral symmetry at low temperture and high density. \\
\indent However, this analysis tells us only second-order phase transition because we evaluate only the RG flow of the four-fermi coupling constant. In other words, we see the behaivor of the curvature of the mesonic effective potential at the origin only, and therefore, we overlook the possibility of the first-order transitions. We should bosonize the four-fermi interactions~\cite{boso1,boso2,boso3} or treat total mass function~\cite{Kuma} in order to evaluate the order of phase transition.

\end{document}